\documentclass[a4paper]{jpconf}
\usepackage{graphicx,epstopdf}
\usepackage{amsmath}
\usepackage{epsfig}
\usepackage{mathdots}
\usepackage{wrapfig}
\newlength{\upit}\upit=0.1truein

\newcommand{\ltappr}{{{\lower4pthbox{$<$} } \atop \widetilde{ \ \ \ }}}
\newlength{\bxwidth}\bxwidth=1.5 truein

\newcommand{\dg}{^{\dagger }}

\newlength{\figwidth}
\newlength{\shift}
\newcommand \bea {\begin{eqnarray} }
\newcommand \eea {\end{eqnarray}}
\newcommand \beg {\begin{equation} }
\newcommand \en {\end{equation}}
\newcommand{\bp}{\mathbf p}

\newcommand{\bk}{{\bf{k}}}

\newcommand{\bsig}{\mbox{\boldmath $\sigma$}}

\begin{document}
\title{Quantum criticality and superconducting pairing in Ce$_{1-x}$Yb$_x$CoIn$_5$ alloys}

\author{Y. P. Singh,$^1$ D. J. Haney,$^1$ I. K. Lum,$^2$ B. D. White,$^2$ M. B. Maple,$^2$ M. Dzero,$^1$ and C. C. Almasan$^1$}
\address{$^1$Department of Physics, Kent State University, Kent, Ohio, 44242, USA, \\
$^2$Department of Physics, University of California at San Diego, La Jolla, CA 92903, USA}                           
\date{\today}

\ead{calmasan@kent.edu}

\begin{abstract}
Charge transport measurements under magnetic field and pressure on Ce$_{1-x}$Yb$_x$CoIn$_5$ single crystalline alloys revealed that:
(i) relatively small Yb substitution suppresses the field induced quantum critical point, with a complete suppression for nominal Yb doping $x>0.20$; (ii) the superconducting transition temperature ($T_c$) and Kondo lattice coherence temperature ($T_{\textrm{coh}}$) decrease with $x$, yet they remain finite over the wide range of Yb concentrations; (iii)  both $T_c$ and $T_{\textrm{coh}}$ increase with pressure; (iv) there are two contributions to resistivity, which show different temperature and pressure dependences, implying that both 
heavy and light quasiparticles contribute to inelastic scattering. We also analyzed the pressure dependence of both $T_{\textrm{coh}}$ and $T_c$ within the composite pairing theory. In the purely static limit, we find that the composite pairing mechanism necessarily causes opposite behaviors of $T_{\textrm{coh}}$ and $T_c$ with pressure: if $T_{\textrm{coh}}$ grows with pressure, $T_c$ must decrease with pressure and vice versa. 
\end{abstract}

\section{Introduction}
Most of the current research efforts in unconventional superconductors are primarily focused on the understanding of their normal state properties, possible symmetries of the superconducting order parameter, as well as the microscopic mechanism of Cooper pairing. Generally, it is believed that superconductivity with $s$-wave symmetry of the order parameter is often realized in materials where 
electron-electron correlations are weak. In contrast, in materials with strong electronic correlations, the superconducting order parameter often develops nodes giving rise to $d$-wave or even $f$-wave pairing symmetries \cite{Norman2011}. It is worth noting that unconventional superconductivity may develop from purely repulsive electron-electron interactions. Although a lot of progress has been made recently in our understanding of unconventional superconductivity in complex materials, the highly correlated nature of the many-body states makes the theoretical and experimental analysis of these materials very challenging. 

The temperature-pressure ($T-P$) and temperature-doping ($T-x$) phase diagrams of the most unconventional superconductors reveal an intricate interplay between magnetism and superconductivity \cite{Pfleiderer2009,Scalapino2012}. Namely, superconductivity emerges from an antiferromagnetic parent state upon doping with an excess of charge carriers, suggesting that the superconducting pairing mechanism could be related to the antiferromagnetic instability of the parent state. Moreover, superconductivity and antiferromagnetism coexist in most unconventional superconductors over a certain space of their $T-P$ and $T-x$ phase diagrams. 
Importantly, there is overwhelming evidence for
an underlying quantum critical point (QCP) separating magnetic and paramagnetic states within the superconducting phase, suggesting that strong magnetic fluctuations play a key role in the emergence of superconductivity. 
\begin{wrapfigure}{l}{0.51\textwidth}
\begin{center}
\includegraphics[scale=0.27,angle=0]{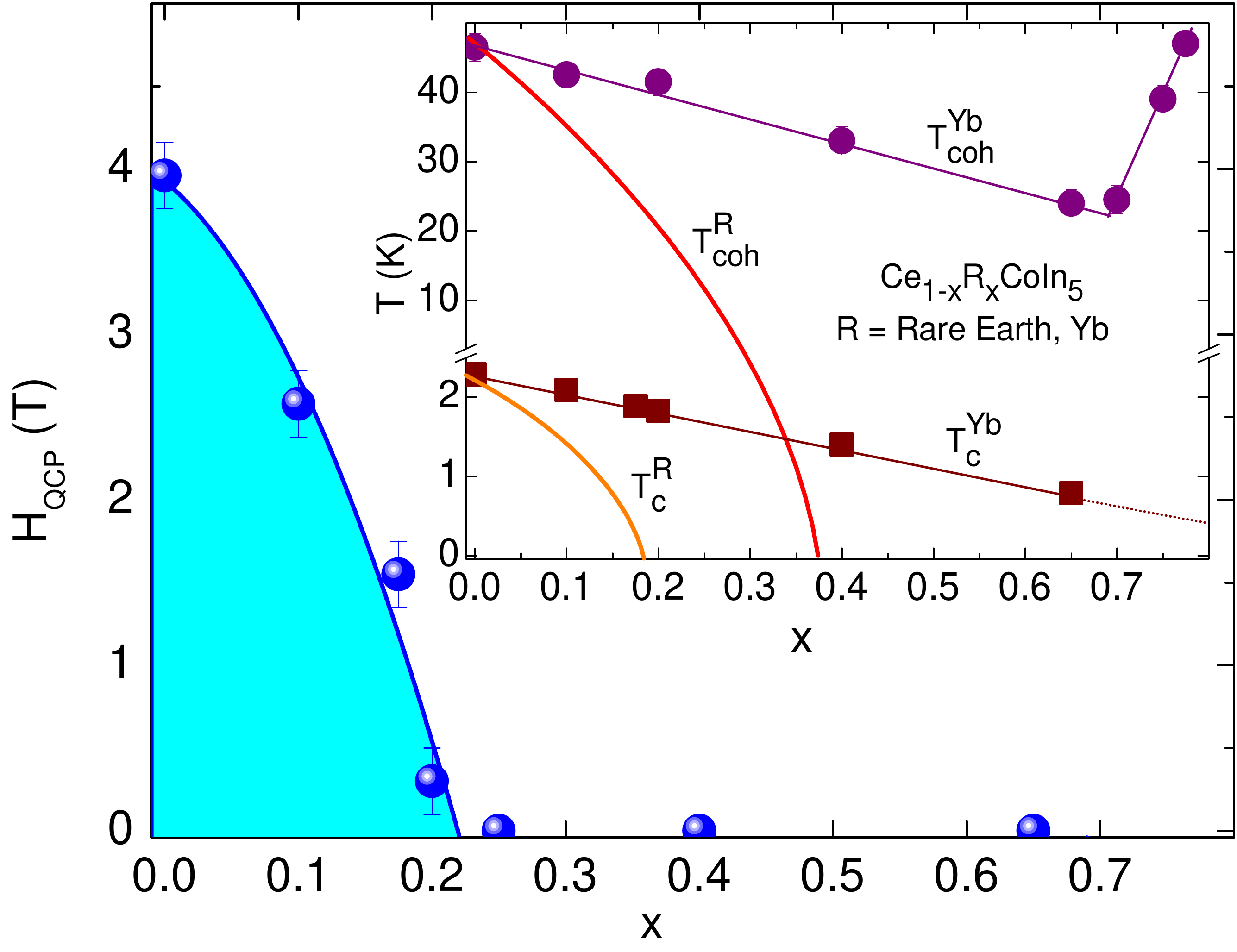}
\end{center}
\caption{\small (Color online) Evolution of the field-induced quantum critical point $H_{QCP}$ of Ce$_{1-x}$Yb$_x$CoIn$_5$ as a function of Yb concentration $x$. Inset: Coherence temperature $T_{\textrm{coh}}$ and superconducting critical temperature $T_c$ for Ce$_{1-x}$R$_x$CoIn$_5$ (superscript R is for rare earth and Yb for ytterbium). The data for the rare earth are taken from Ref. \cite{Paglione2007}.}
\label{Phase Diagram}
\end{wrapfigure}

Although there are many examples where unconventional Cooper pairing is driven by the system's proximity to a magnetic QCP and is therefore mediated by an exchange of 
paramagnetic fluctuations, there are also a few notable exceptions. For instance, no signatures of magnetic fluctuations are found in the heavy-fermion superconductors PuCoGa$_5$ and PuRhGa$_5$ \cite{Curro2005,Sakai2005,Bauer2012}, most probably due to a mixed-valence state of the Pu ion. Similarly, in recently discovered high-temperature heavy-fermion superconductor Np$_2$Pd$_5$Al$_2$, magnetic susceptibility has a Curie-Weiss temperature dependence down to $T_c$, signaling the absence of pronounced magnetic interactions between the Np moments \cite{Aoki2007}. Another example is the Fe-based superconductor LiFeAs \cite{Borisenko2010}. Intriguingly, although there are multiple evidences for the presence of strong magnetic fluctuations in CeCoIn$_5$ \cite{Hu2012}, superconductivity remains robust with respect to 
alloying this parent compound by Yb substitution on the Ce site \cite{Shu2011}. This robustness of superconductivity against substitution-induced disorder points towards an alternative microscopic origin of superconductivity in this system. 

In this paper, we focus on the the nature of the pairing mechanism and the issue of quantum criticality in Ce$_{1-x}$Yb$_x$CoIn$_5$ alloys. We also present a  theoretical study of the pressure dependence of the Kondo lattice coherence and superconducting critical temperatures within the frame of the composite pairing theory. 

\section{Transport properties}
\paragraph{\bf General remarks.}
The resistivity of Ce$_{1-x}$Yb$_x$CoIn$_5$ ($0.00 \leq x \leq 0.75$) alloys exhibits properties typical of heavy fermion systems \cite{Shu2011,Hu2013,Singh2014}, but our detailed analysis of charge transport measurements in the presence of magnetic field and hydrostatic pressure have revealed many interesting and unusual features that emerge with Yb doping. These results allowed us to extract information about the evolution of the magnetic-field-tuned QCP present in CeCoIn$_5$ \cite{Paglione2003} with Yb doping and about the nature of the superconducting pairing.

At ambient pressure, the parent compound CeCoIn$_5$ of the Ce$_{1-x}$R$_x$CoIn$_5$ series (R is a rare earth) has been shown to be  in close proximity to an antiferromagnetic QCP that is at an inaccessible negative pressure $P_c$ \cite{Sidorov2002,Nair2010}. With magnetic field ($H$) and $P$ as control parameters, thermal expansion \cite{Zaum2011} and current-voltage measurements in the mixed state \cite{Hu2012} have revealed a quantum critical line in the  $H$-$P$ plane of CeCoIn$_5$ at $T=0$ K. In addition, magneto-transport measurements show that Kondo coherent scattering dominates the physics of CeCoIn$_5$ at relatively high temperatures \cite{Hu2013,Singh2014}. Substitutional disorder by alloying the $f$-electron sites with magnetic or non-magnetic rare earth ions has been employed as a tuning parameter. Irrespective of the magnetic nature of the substituent, the response of all these compounds is the same - the suppression of both superconducting transition temperature ($T_c$) and Kondo lattice coherence ($T_{\textrm{coh}}$) temperatures with substitutional disorder - with a full suppression of  $T_c$ at about $20 \%$ and  $T_{\textrm{coh}}$  at around $40\%$ of substitutional disorder (see inset to Fig. \ref{Phase Diagram}) \cite{Paglione2007}. 

\paragraph{\bf Experimental details.} Single crystals of Ce$_{1-x}$Yb$_x$CoIn$_5$ ($0\leq x \leq 0.75$), where $x$ is the nominal Yb doping, were grown using an indium self-flux method. The quality and structural details of the grown crystals were checked with X-ray powder diffraction and energy dispersive X-ray techniques. The actual Yb doping differ from the nominal concentration and a detailed analysis of the relationship between actual and nominal doping has been discussed in Ref. \cite{Jang2014}. Through out this article we will use the nominal concentration to be able to discuss our results in the context of other published fundings on this system.

The single crystals studied have a typical size of $1.0 \times 0.5 \times 0.1$  mm$^3$, with the $c$-axis along the shortest dimension of the crystals. The crystals were etched in concentrated HCl to remove the indium left on the surface during the growth process and were then rinsed thoroughly in ethanol. Four leads were attached to the single crystals, with the electrical current $I \parallel a$-crystallographic axis. Charge transport measurements were performed for temperatures between 2 K and 300 K, an applied magnetic field up to 14 T, and hydrostatic pressures up to 8.6 kbar. We also performed transverse ($H \perp I$) in-plane ($I \parallel a$-axis) magnetoresistance (MR) measurements, where MR is defined as $\Delta\rho_a^{\perp}/\rho_a=(\rho_a^{\perp}-\rho_a(0))/\rho_a(0)$.

\begin{wrapfigure}{l}{0.51\textwidth}
\begin{center}
\includegraphics[scale=0.27,angle=0]{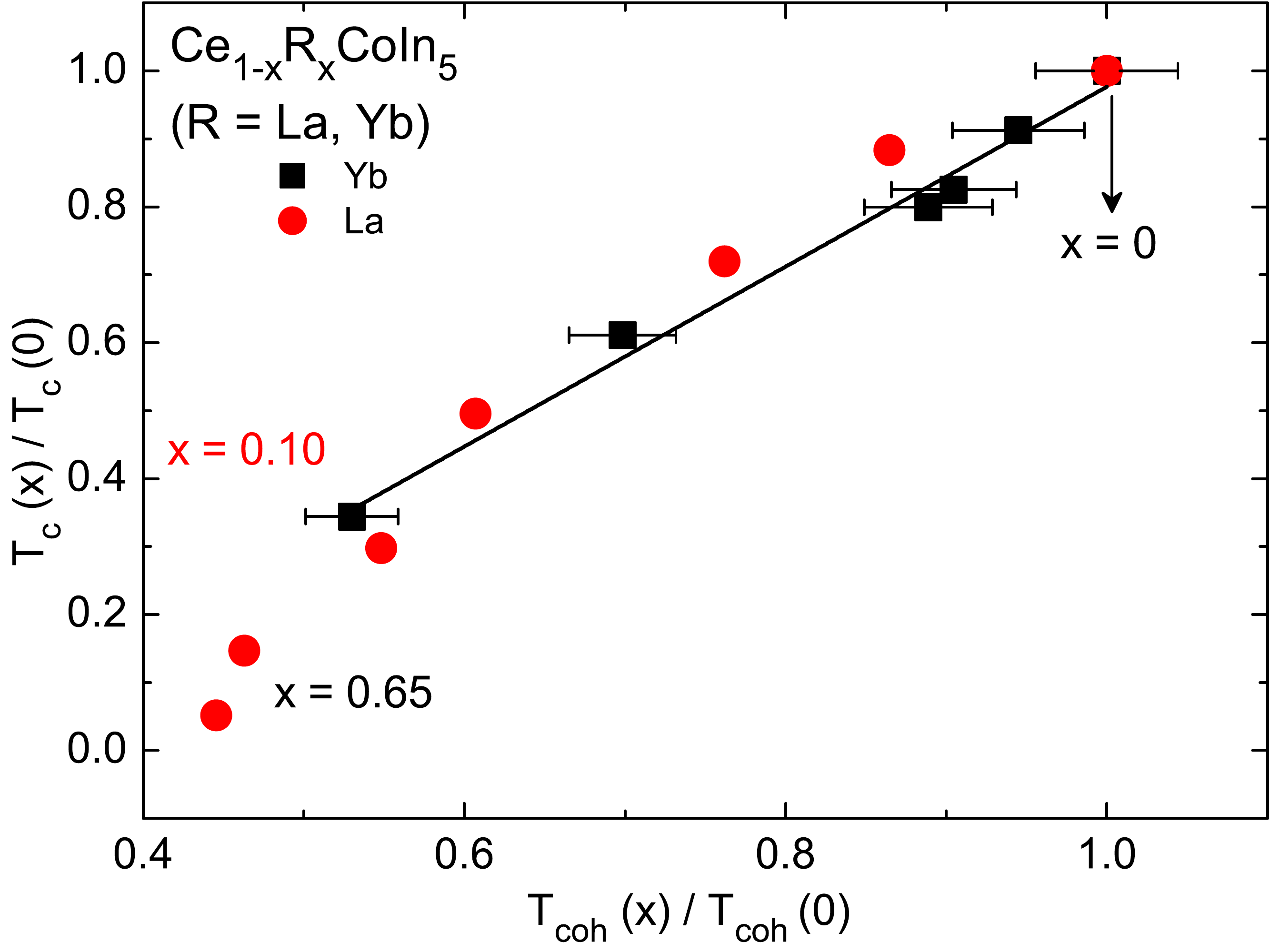}
\end{center}
\caption{\small (Color online) Plot of superconducting critical temperature $T_c$ vs. Kondo lattice coherence temperature $T_{\textrm{coh}}$ normalized by the corresponding values of the undoped sample, for Yb and La \cite{Petrovic2002} doping on Ce-site.}
\label{Scaling}
\end{wrapfigure}

\paragraph{\bf Suppression of the field induced QCP.} Major deviations from standard behavior \cite{Paglione2007} are observed when the rare earth is in the intermediate-valence state as is Yb in Ce$_{1-x}$Yb$_x$CoIn$_5$ \cite{Shu2011,Dudy2013}: superconductivity and Kondo coherence are weakly suppressed with Yb doping and extend to large nominal Yb concentrations (inset to Fig.~\ref{Phase Diagram}), in contrast with the behavior of all the other rare-earth substitutions discussed above, and, in addition, these alloys display a non-Fermi liquid (NFL) behavior for the whole Yb doping range \cite{Shu2011,Hu2013,Kim2014}. The fact that $T_c$ and $T_{\textrm{coh}}$ are unusually robust to Yb substitution suggests that the strong pair-breaking effects of impurity substitution are reduced by the cooperative intermediate valence state of Yb \cite{Shu2011,Dudy2013}.  In fact, it has  been proposed that there are strong impurity correlations between Yb ions at low Yb concentrations that result in a healing effect on $T_c$ and $T_{\textrm{coh}}$ \cite{Dzero2012}. On the other hand, the field-induced quantum critical point ($H_{QCP}$) is finite for smaller doping, but it is suppressed rapidly to almost zero at about $20\%$ of nominal Yb doping (Fig. \ref{Phase Diagram}), indicating that this doping is close to the critical value $x_c$. Recent penetration depth measurements suggest the appearance of a nodeless superconducting order parameter state at $20\%$ nominal Yb doping \cite{Kim2014}.  The significant change in Fermi-surface topology revealed by these penetration depth and de Haas-van Alphen measurements \cite{Polyakov2012} confirms that, indeed,  the alloy with $20\%$ nominal Yb doping is quantum critical. 

The rapid suppression of $H_{QCP}$ with Yb doping, along with the robustness of $T_{\textrm{coh}}$ and $T_c$  also suggests that spin fluctuations are not the ``glue" for Cooper pairs. In addition, the scaling of the doping dependent $T_c$ and $T_{\textrm{coh}}$, normalized to the corresponding values for CeCoIn$_5$ (Fig. ~\ref{Scaling}), shows that many-body coherence and superconductivity have the same microscopic origin: the hybridization between the conduction electrons with the localized Ce $f$-moments. The data for La substitution on the Ce site, also shown in Fig.~\ref{Scaling}, reveal that the same scaling works up to $10\%$ of La doping, which is an indication that the onset of many-body coherence in the Kondo lattice and the emergence of the superconductivity have the same physical origin in this system too. Thus, we conclude that the Cooper pair formation in both La- and Yb-doped CeCoIn$_5$ occurs on the heavy Fermi surface. Furthermore, the correlations among Yb ions, which are governed by its intermediate valence state, significantly reduce pair breaking due to the disorder induced by Yb substitution. 


\begin{figure}
\centering
\includegraphics[width=0.75\linewidth]{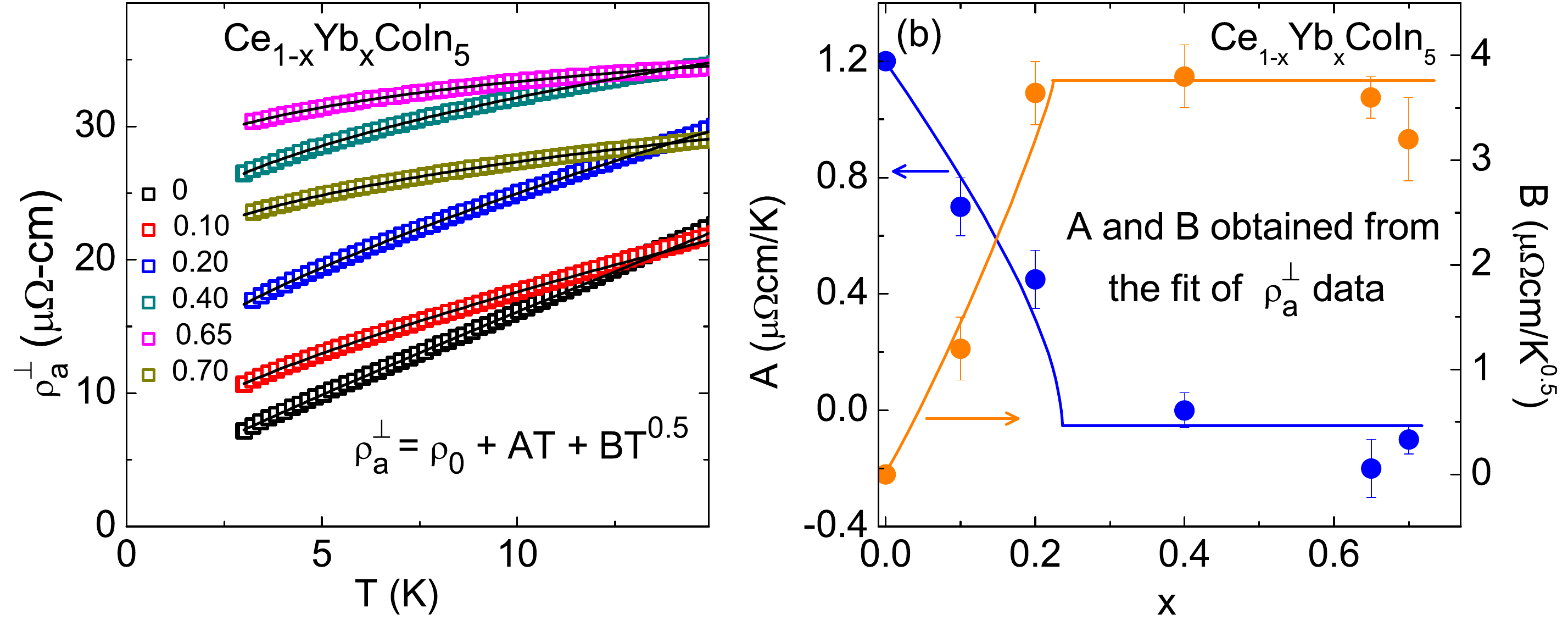}
\caption{\small (Color online)(a) Fits of the resistivity data with $\rho_a^{\perp} = \rho_0 + AT+B \sqrt T$ for different doping levels of Yb in the temperature range 4 K $\leq T\leq$ 15 K. (b) Doping $x$ dependence of parameters $A$ and $B$  obtained from fits of the resistivity data shown in panel (a).} 
\label{Resistivity}
\end{figure} 

\paragraph{\bf Temperature dependence of resistivity.}
Another interesting phenomenon that is consistently observed in Ce$_{1-x}$Yb$_x$CoIn$_5$ alloys over the whole Yb doping range is the sub-linear $T$  dependence of resistivity in the normal state, just above $T_c$  \cite{Shu2011}. This is quite puzzling since the presence of the NFL behavior is usually associated with the presence of a QCP that, nevertheless, is fully suppressed in this system for $x \approx 0.20$. Thus, the NFL behavior at higher Yb-doping can not be attributed to the presence of quantum fluctuations. With this in mind, we further investigated the temperature dependence of in-plane resistivity with $H||c$-axis ($\rho_a^{\perp}$). Our analysis has revealed that $\rho_a^{\perp}$ has a $\sqrt{T}$ dependence, except for the lower doping levels ($x\leq$ 0.20) where it exhibits an additional linear-in-$T$ contribution; thus, we were able to fit all the data very well for $3 <T<$ 15 K [see Fig.~\ref{Resistivity}(a)] with
\begin{equation}\label{resistivity}
\rho_a^{\perp}(x,T) = \rho_0 + AT+B\sqrt T,
\end{equation} 
where $\rho_0$, $A$, and $B$ are doping- and field-dependent fitting parameters. Figure~\ref{Resistivity}(b) shows $A$ and $B$ vs doping. A few features of this graph are worth noting. First, $A$ decreases rapidly with increasing $x$ and is negligible for $x>$ 0.20. This observation is not surprising because the QCP is also suppressed for $x>$ 0.20, and alloys with higher Yb doping do not show any sign of quantum fluctuations. Thus, as expected, there is a direct correlation between the linear-in-$T$ contribution in resistivity and the presence the field-induced QCP. Secondly, the $\sqrt{T}$ contribution to resistivity is absent for the stoichiometric parent compound, i.e., $B(x=0)=0$. Furthermore, this contribution initially increases with increasing Yb concentration up to $x\approx$ 0.20 and then it saturates for higher Yb doping. The microscopic origin may be due to changes in the electronic structure of Yb:
it has been found that Yb ion changes its valence state at $x\approx 0.20$, from a magnetic-valence state at low $x$ to an intermediate-valence state at high $x$ \cite{Dudy2013}.
\begin{figure}
  \begin{minipage}[c]{0.48\textwidth}
    \includegraphics[width=\textwidth]{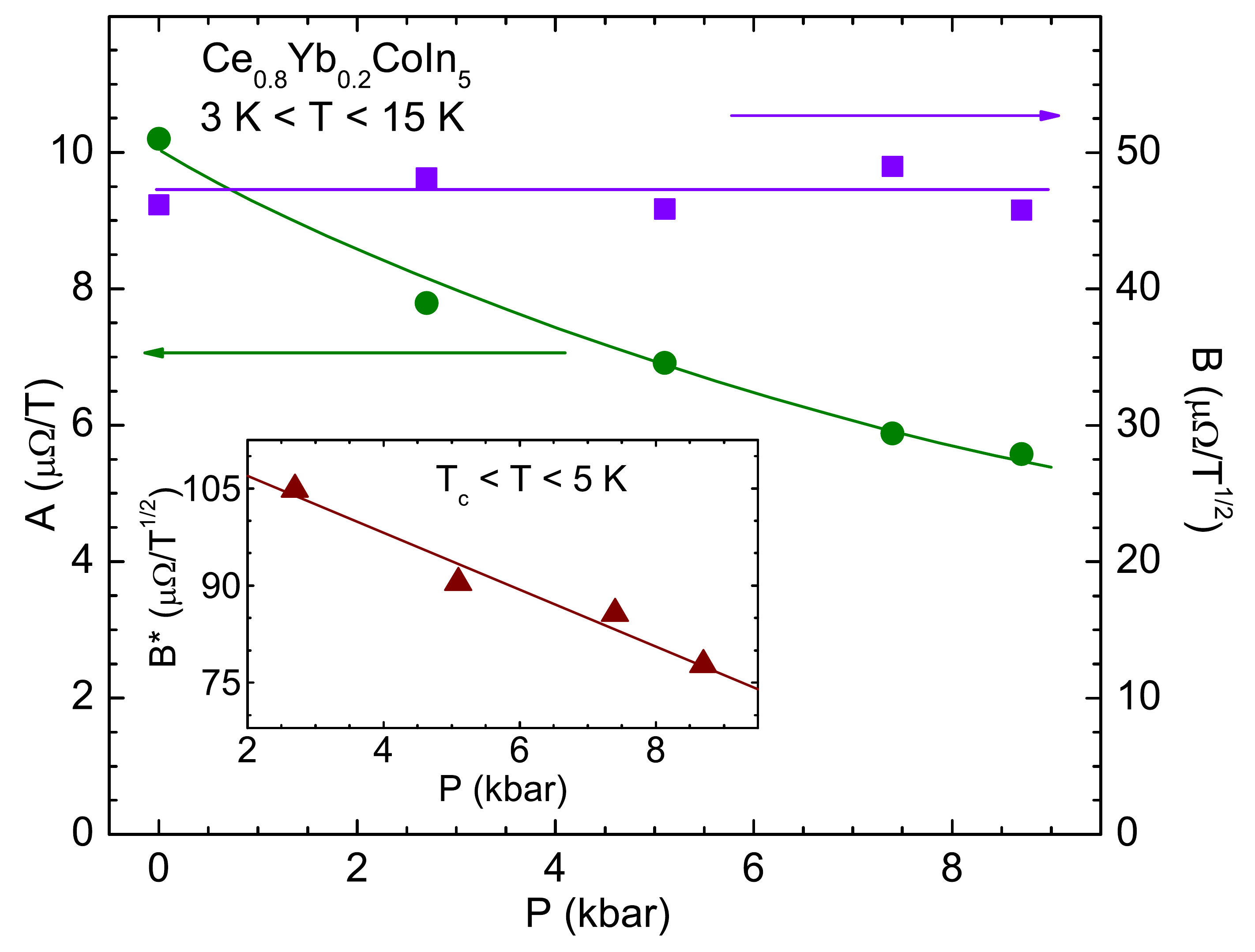}
  \end{minipage}\hfill
  \begin{minipage}[c]{0.38\textwidth}
    \caption{(Color online) Pressure $P$ dependence of parameters $A$ and $B$ obtained from the fits of the resistivity data for Ce$_{0.8}$Yb$_{0.2}$CoIn$_5$ with $\rho_a^{\perp}(P,T) = \rho_0 + AT+B \sqrt T$. The fitting is performed over the temperature range 3 K $\leq T\leq$ 15 K. Inset:  $P$ dependence of parameter $B^*$ obtained from the fits of the resistivity data with $\rho_a^{\perp}(H) = \rho_0+B^* \sqrt T$.} 
 \label{PressureA}
  \end{minipage}
\end{figure}

\begin{figure}
  \begin{minipage}[c]{0.48\textwidth}
    \includegraphics[width=\textwidth]{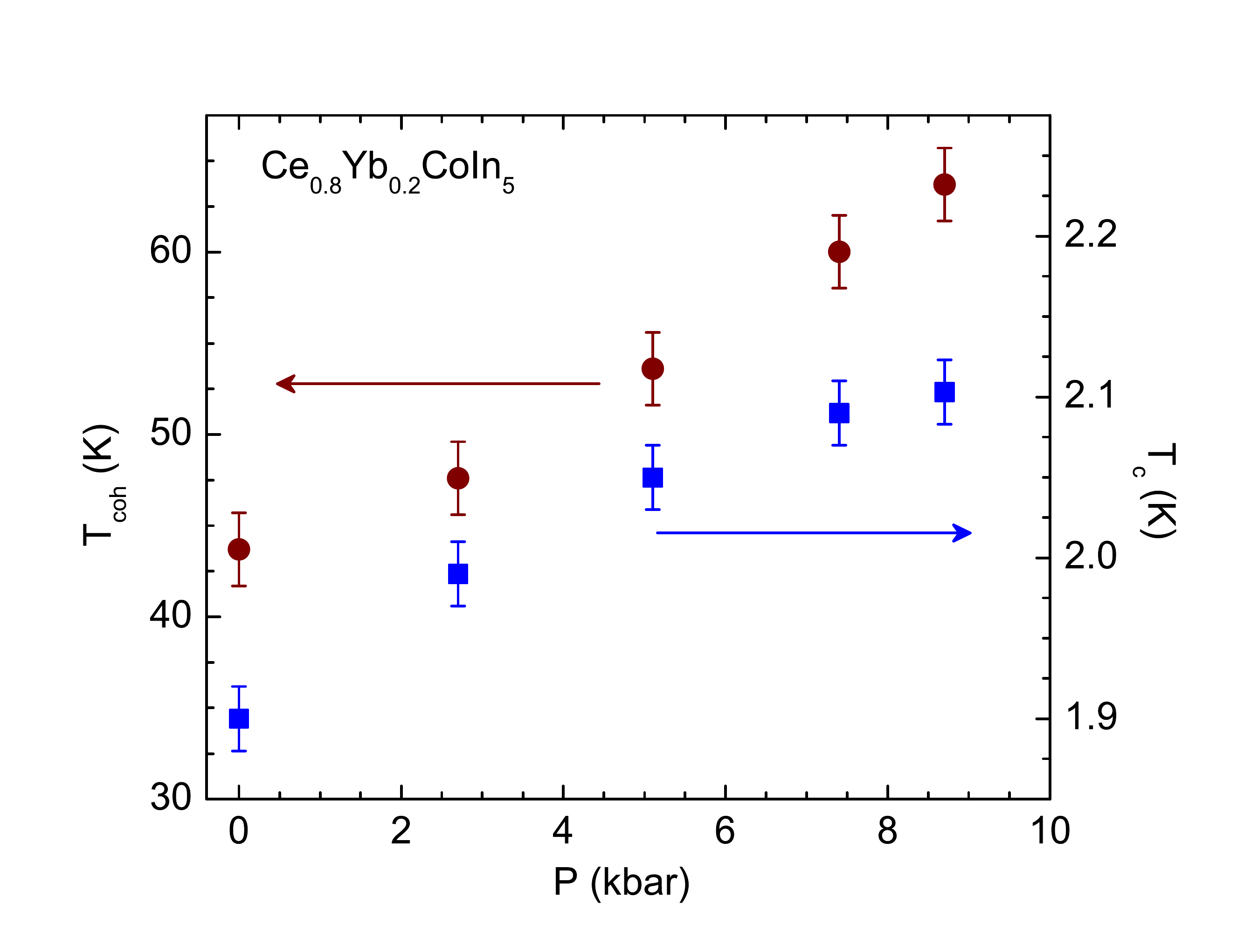}
  \end{minipage}\hfill
  \begin{minipage}[c]{0.38\textwidth}
    \caption{(Color online) Pressure $P$ dependence of coherence temperature $T_{\textrm{coh}}$ and superconducting transition temperature $T_c$ for Ce$_{0.8}$Yb$_{0.2}$CoIn$_5$ for pressures up to 8.5 kbar.} \label{PressureB}
  \end{minipage}
\end{figure}
\paragraph{\bf Charge transport under pressure: heavy and light Fermi surfaces.}
To further explore the origin of the NFL behavior in Ce$_{1-x}$Yb$_{x}$CoIn$_5$ alloys and the possibility of an alternative pairing mechanism in this system, we performed resistivity measurements under hydrostatic pressure on the quantum critical alloy, i.e., on the $x= 0.20$ crystals. These measurements have revealed that the resistivity data are best fitted with
$\rho_a^{\perp}(P,T) = \rho_{0} + AT+B\sqrt T$ for 3 K $< T <$ 15 K
and 
$\rho_a^{\perp}(P,T) =  \rho_{0}+B^*\sqrt T$ for $T_c < T < 5$ K.
The main panel in Fig.~\ref{PressureA} shows that $A$ decreases with increasing $P$, which is expected because it is attributed to quantum spin fluctuations, which are suppressed with pressure in Ce based heavy-fermion alloys \cite{Jaccard1999,Grosche1996,Hu2012,Almasan2014}. On the other hand, $B$ is insensitive to pressure. We have shown in Fig.~\ref{Resistivity}(b) that the $\sqrt{T}$ contribution is absent in the parent compound CeCoIn$_5$, but it increases with increasing doping. This behavior, together with the fact that $B$ is independent of pressure, suggests that the origin of the $\sqrt{T}$ contribution for 3 $< T <$ 15 K is inelastic scattering of quasiparticles from the small Fermi surface. On the other hand, the absence of any linear-in-$T$ resistivity just above $T_c$ along with the decrease of $B^*$ with increasing $P$ (inset to Fig.~\ref{PressureA}) show that superconducting fluctuations dominate this $T$ region and that the inelastic scattering events leading to the $\sqrt T$ dependence in this temperature range involve quasiparticles from the heavy Fermi surface, respectively. This latter result is consistent with scattering of composite pairs in this material. Further discussion about composite pairing is provided in the theoretical section below. 

Finally, Fig.~5 shows that both $T_{\textrm{coh}}$ and $T_c$ increase with increasing pressure up to 8.5 kbar.  As discussed in the theory section below, this result is inconsistent with the composite pairing mechanism, which suggests an increase in $T_{\textrm{coh}}$ and a decrease in $T_c$ with increasing pressure. Nevertheless, as discussed below, considerations of the effect of quantum valence fluctuations could reconcile the composite pairing theory and these experimental results. 

\section{Pressure Effects in Composite Pairing Superconductivity}
Composite pairing theory has recently emerged as a prominent microscopic mechanism for superconducting pairing
in heavy-fermion materials \cite{Coleman1999,Flint2008,Flint2010,Flint2011}. At the heart of the theory is the idea 
that virtual fluctuations of an $f$-electron ion between magnetic (say $f^1$) and non-magnetic ($f^0$ and $f^2$)
valence states become resonant \cite{Cox1997} and, in principle, can promote superconducting pairing \cite{Coleman1999}. It is crucial that the emerging superconducting amplitude is anisotropic in momentum, signaling an onset of unconventional superconductivity. This in turn implies that the composite pairing can only be realized in the lattice of magnetic moments. It is also worth mentioning that, initially, the idea of the composite pairing has been developed in the context of odd-frequency Cooper pairing and its realization in heavy-fermion materials \cite{Coleman1993,Coleman1994,Hoshino2014}. Subsequently, it was realized that even-frequency composite pairing can be regarded as an alternative to odd-frequency pairing with the order parameter given by the expectation value, which also contains a local spin operator \cite{Hoshino2014}. 
\begin{wrapfigure}{l}{0.51\textwidth}
\begin{center}
\includegraphics[scale=0.2,angle=0]{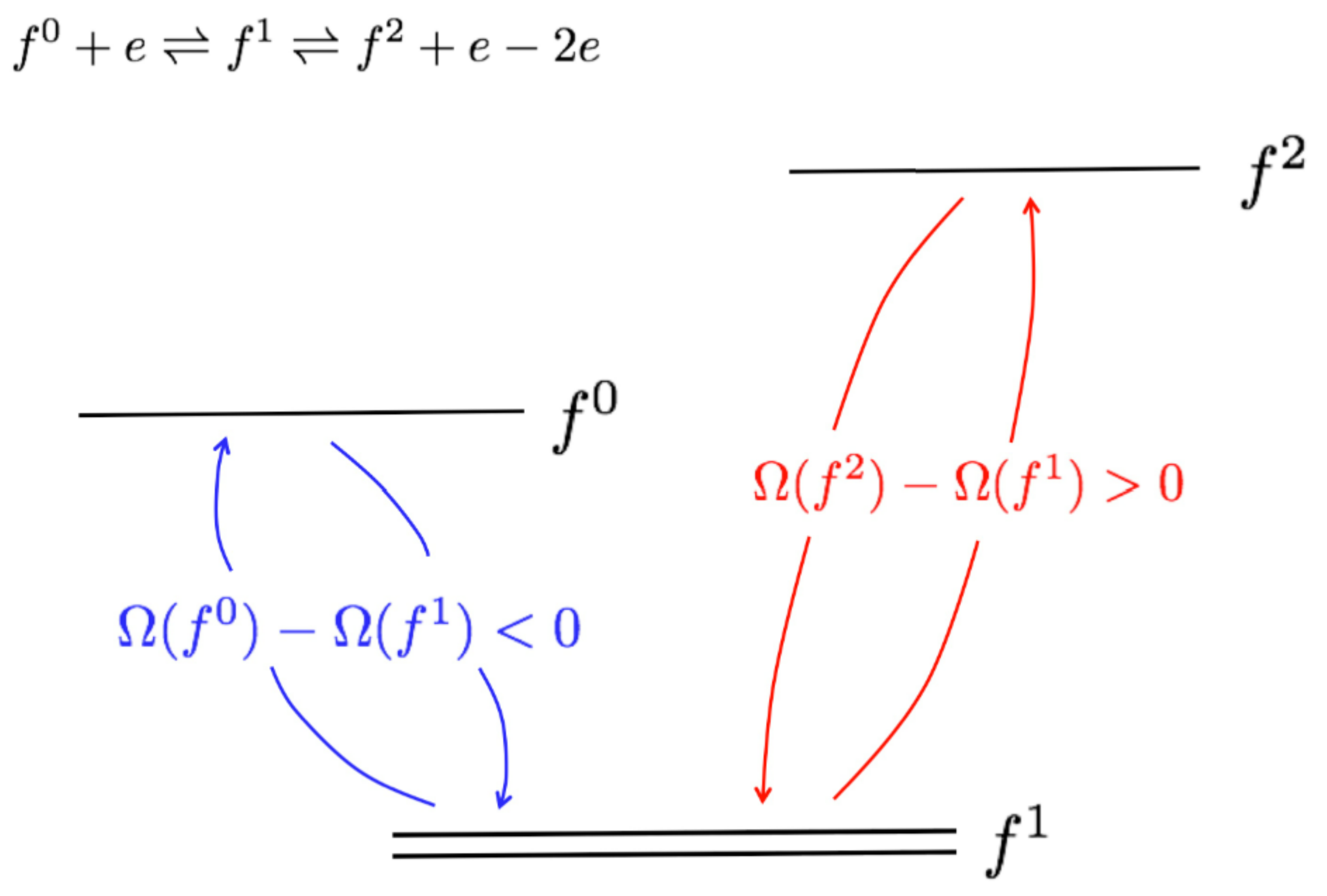}
\end{center}
\caption{\small Schematic presentation for the virtual charge fluctuations between magnetic $f^1$ and
non-magnetic valence configurations: $f^1\rightleftharpoons f^0+e$ via the first conduction channel and 
$f^1\rightleftharpoons f^2-e$ via the second conduction channel. Since the change in the ionic radius is opposite 
for the two channels, we expect the opposite behavior for the change in the corresponding coherence temperatures with pressure.}
\label{Fig6}
\end{wrapfigure}

In this Section we theoretically study the effect of hydrostatic pressure on composite pairing. Specifically, 
we evaluate the pressure dependence of $T_{\textrm{coh}}$ and $T_c$. For simplicity we will ignore the presence of disorder, so strictly speaking for the arguments presented below we assume the presence of a spatially homogeneous lattice. Our primary goal here is to show that the microscopic mechanism for the composite pairing necessarily implies opposite behaviors of $T_{\textrm{coh}}$ and $T_c$ with pressure, which can be traced to the opposite changes in the ionic sizes as the resonance valence fluctuations take place. 

\subsection{Model}\label{ModelLargeN}
We consider the two-channel Kondo lattice Hamiltonian, which is obtained from the Anderson lattice model by formally integrating out high-energy states by means of the Schrieffer-W\"{o}lff transformation \cite{Cox1997,Coleman1999,Flint2008}. We have:
\begin{equation}\label{H2ch}
H= \sum_{\bk \sigma }\epsilon_{\bk}c_{\bk \sigma}\dg
c_{\bk \sigma}+
{\textstyle \frac{1}{2}}
\sum_{\bk\alpha,\bp\beta}\sum\limits_{j}J_{\bk ,\bp} c\dg_{\bk \alpha }
 c_{\bp\beta }\left({\vec \sigma}_{\alpha\beta}\cdot{\vec S}_{f} (j)\right) e^{i (\bp-\bk
 )\cdot {\bf R}_{j}},
\end{equation}
where $c_{\bk \sigma}\dg$ is a fermionic creation operator, $\bk$ is a momentum, $\sigma=\uparrow,\downarrow$, 
$\epsilon_\bk=-\frac{t}{4}(\cos k_x+\cos k_y)-\mu$,
$t$ is a hopping amplitude, $\mu$ is a chemical potential, and ${\vec S}_f(j)=\frac{1}{2}f_{j\nu}\dg {\vec \sigma}_{\nu\eta}f_{j\eta}$,
written using the Abrikosov fermionic representation, accounts for the localized cerium $f$-moments at site ${\vec R}_j$. Note that, for simplicity, we choose the two-dimensional spectrum for the conduction electrons. The momentum dependent exchange couplings $J_{\bp,\bk}= J_{1}\phi_{1\bk }\phi_{2\bp}+J_{2}\phi_{2\bk }\phi_{2\bp}$
include exchange couplings $J_{1,2}>0$ for the electrons in the first and second conduction channels, while
$\phi_{\Gamma\bk}$ are the corresponding form-factors. Without loss of generality, we choose them in the following form $\phi_{1\bk}=1$ ($s$-wave) and  $\phi_{2\bk}=\cos k_x-\cos k_y$ ($d$-wave).
The mean-field analysis of the model (\ref{H2ch}) shows that at $T_{\textrm{coh}}$ the heavy-fermion state
forms. Within the mean-field approximation, the formation of the heavy-fermion state is governed by the development
of the non-zero expectation value $\langle c_{j\alpha}\dg f_{j\alpha}\rangle$ at each site. Interestingly, for $J_2<J_1$ 
an anomalous expectation value can develop, signaling the onset of either a charge-density wave state or superconductivity.
The proper choice of the phase stabilizes the superconducting state with a critical temperature \cite{Coleman1999}
$T_c\sim T_{coh}\exp\left(-1/\nu_FJ_2\right)$ where $\nu_F$ is the density of states at the chemical potential. This expression is reminiscent of the BCS weak-coupling
expression for the critical temperature. As we will show below, $T_c$ shows a strong pressure dependence. Finally, we note that the mean-field theory results at ambient pressure are reproduced by various numerical approaches \cite{Anders2002PRB,Anders2002EPJ,Hoshino2014}. More importantly, the composite pairing mechanism for superconductivity can be extended to systems with mixed-valence \cite{Flint2011}. However, to this date, the mean-field
theory of the composite pairing state for the $f$-electron systems in the mixed-valence regime has not been numerically confirmed.

\subsection{Mean-field theory: effect of hydrostatic pressure}
The mean-field theory is formulated by performing the decoupling in the interacting part of the Hamiltonian (\ref{H2ch}).
This is an approximation which becomes exact in the limit when the number of fermionic flavors $N$ goes to infinity.
Therefore, to make our mean-field approximation controlled,  we generalize our model from SU(2) to symplectic-$N$ \cite{Flint2008} by replacing the Pauli spin operators $\bsig_{\alpha \beta
}\rightarrow ( {\bsig}_{N})_{\alpha \beta }$. At ambient pressure we find:
\beg
{\cal F}_0
=-NT\sum\limits_{\bk
,\pm}\log[2\cosh(\beta\omega_{\bk\pm}/2)]+ N{\cal N}_{s}\sum_{\Gamma= 1,2}\frac{v_\Gamma^2}{J_\Gamma},
\en
where $\beta=1/k_BT$, ${\cal N}_s$ is a number of lattice sites, $v_{1,2}$ are corresponding mean-field amplitudes that describe the onset of coherence and superconductivity, while $\omega_{\bk\pm}$ account for dispersion of newly formed
electron bands $\omega_{\bk\pm}=
\sqrt{\alpha_\bk\pm(\alpha_\bk^2-\gamma_\bk^2)^{1/2}}$, 
where we have introduced functions $\alpha_\bk=v_{\bk +}^2+\frac{1}{2}\left(\epsilon_\bk^2+\lambda^2\right)$, 
$\gamma_\bk^2=(\epsilon_\bk\lambda-v_{\bk -}^2)^2+4 (v_{1\bk}v_{2\bk})^2$,
$v_{1\bk }= v_{1}\phi_{1\bk }$, $v_{2\bk }= v _{2}\phi_{2\bk }$, and $v_{\bk \pm}^{2}=v_{1\bk}^2\pm v_{2\bk}^2$.

For the Ce ions, the change in the $f$-shell occupation is positive due to its electronic nature, so that the leading resonance scattering involves conduction electrons in the first channel and a zero-energy boson with amplitude $v_1$, and an electron: $f^{1}(j,m)\rightleftharpoons f^{0}(j,m)+e$.
In contrast, the resonance scattering in the second conduction channel involves a zero energy boson with an amplitude $v_2$ and a hole: $f^{1}(j,m)\rightleftharpoons f^{2}(j,m)+e-2e$. When resonance develops in both channels, for the total volume of the system within the mean-field theory, we write: \cite{Zhang2002}
\beg\label{eq:totvol}
\begin{split}
\Omega_t=\Omega(f^0)v_1^2+(1-v_1^2-v_2^2)\Omega(f^1)+\Omega(f^2)v_2^2,
\end{split}
\en
where $\Omega(f^n)$ are the cell volumes for the singlet ($n=0,2$) states and a doublet ($n=1$) state.
\begin{figure}[h!]
\includegraphics[scale=0.25,angle=0]{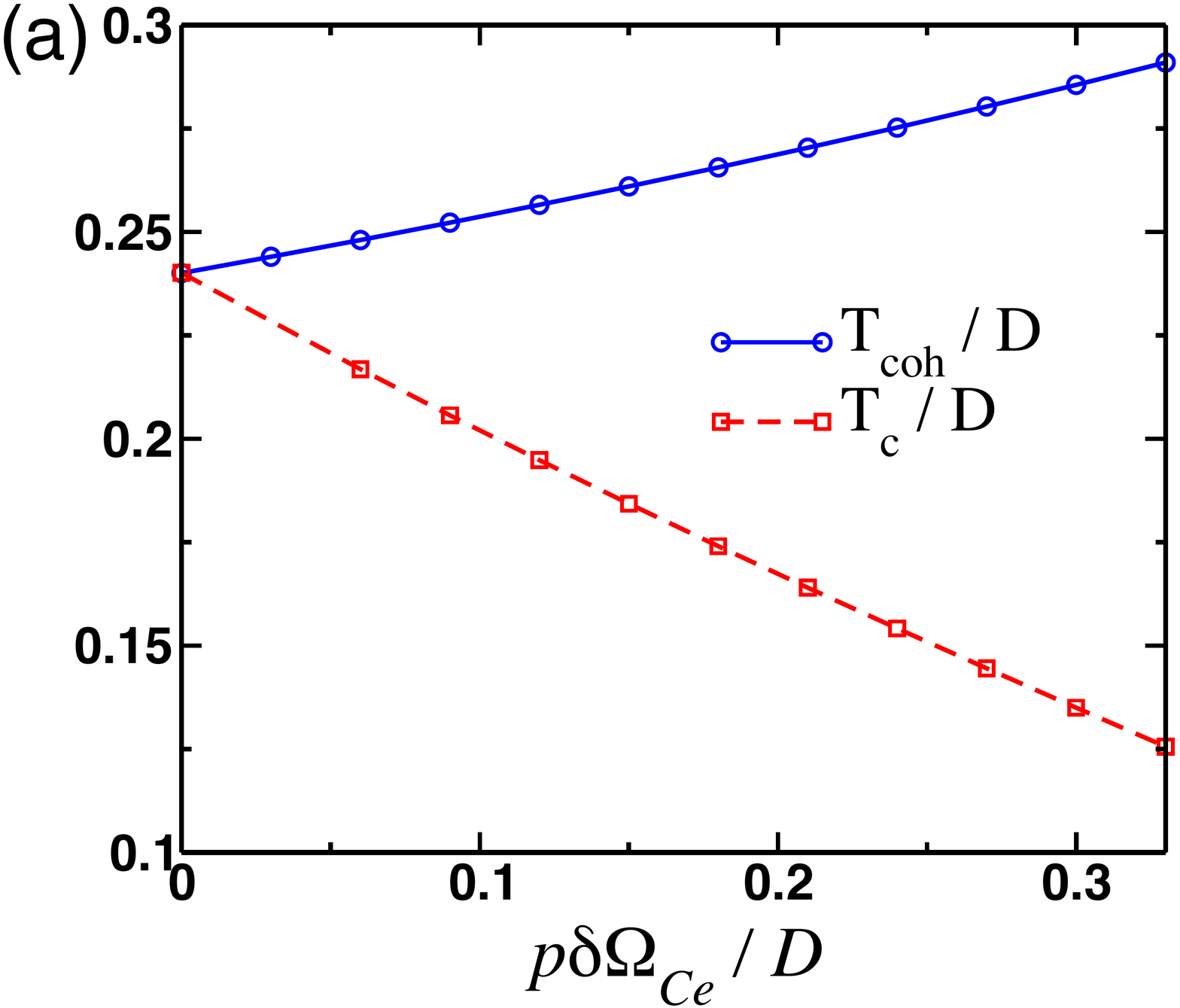}
~\includegraphics[scale=0.25,angle=0]{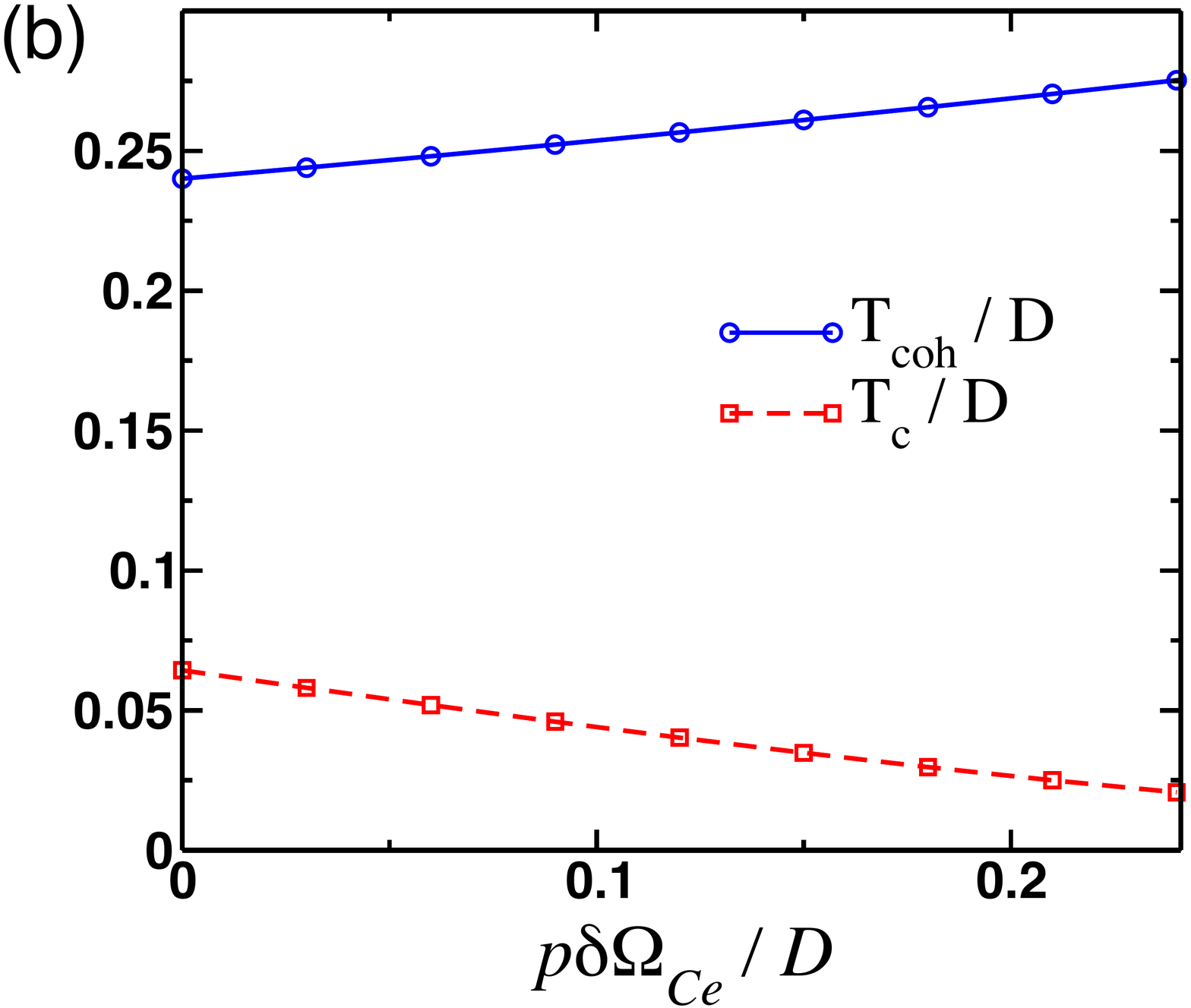}
\caption{\small (Color online) Plots of the pressure dependence of the Kondo lattice coherence temperature $T_{\textrm{coh}}$ 
and superconducting critical temperature $T_c$ from the solution of the mean-field theory for the two-channel 
Kondo lattice model. The first screening channel corresponds to the fluctuations between $f^1$ and $f^0$ cerium valence states, so that the change in the ionic volume is negative $\delta\Omega_{1\textrm{ch}}=\Omega(f^0)-\Omega(f^1)<0$. 
The amplitude $v_1$ becomes non-zero at $T=T_{\textrm{coh}}$ as the resonance $f^1\rightleftharpoons f^0+e$ develops. Similarly, the superconductivity is driven by the hybridization of the conduction electrons with the $f$-states in the second channel corresponding to the fluctuations between $f^1$ and doubly occupied singlet $f^2$: $f^1\rightleftharpoons f^2+e-2e$. Since the $f^2$ state has a larger ionic volume $\delta\Omega_{2\textrm{ch}}=\Omega(f^2)-\Omega(f^1)>0$. Here we define $\delta\Omega_{Ce}=-\delta\Omega_{\textrm{1ch}}=\delta\Omega_{2\textrm{ch}}$.
We use the following set of parameters: $\mu=-0.125t$, $D=t$ and $J_1=0.5t$. Panel (a): $J_2=1.05J_1$. Panel (b): 
$J_2=0.7J_1$.}
\label{TcTcohP}
\end{figure}
It is convenient to introduce the change in the cell volumes: $\delta\Omega_{\textrm{1ch}}=\Omega(f^0)-\Omega(f^1)$
is negative and accounts for the difference in cell volumes between two $f$-ion configurations for the resonance in the first channel. Similarly, $\delta\Omega_{\textrm{2ch}}=\Omega(f^2)-\Omega(f^1)$ is positive and yields the difference in volume between ionic configurations for the resonance in the second channel, Fig.~\ref{Fig6}. In what follows, we assume 
that $\delta\Omega_{\textrm{2ch}}\approx-\delta\Omega_{\textrm{1ch}}=\delta\Omega_{\textrm{Ce}}$. For the free energy we find:
\beg\label{FreeEnergy}
{\cal F}={\cal F}_0+P\Omega_t.
\en

The mean-field equations are obtained by minimizing the free energy (\ref{FreeEnergy}) with respect to $\lambda$ and $(v_{\Gamma})^{2}$ ($\Gamma = 1,2$). This yields:
\begin{equation}
\begin{split}
&\frac{1}{{\cal N}_s}\sum\limits_{\bk\pm}\frac{\tanh(\omega_{\bk\pm}/{2T})}{2\omega_{\bk\pm}}
\left(\lambda\pm\frac{\lambda\alpha_\bk-\epsilon_\bk(
\epsilon_\bk\lambda-v_{\bk -}^2)}{\sqrt{\alpha_\bk^2-\gamma_\bk^2}}\right)=0,\\
&\frac{1}{{\cal N}_s}\sum\limits_{\bk\pm}\phi_{1{\bk}}^2\frac{\tanh(\omega_{\bk\pm}/{2T})}{2\omega_{\bk\pm}}
\left(2\pm
\frac{
(\epsilon_\bk+\lambda)^2
}
{\sqrt{\alpha_\bk^2-\gamma_\bk^2}}
\right)=\frac{4}{J_1}+4P\delta\Omega_{\textrm{1ch}}, \\
&\frac{1}{{\cal N}_s}\sum\limits_{\bk\pm}\phi_{2{\bk}}^2\frac{\tanh(\omega_{\bk\pm}/{2T})}{2\omega_{\bk\pm}}
\left(2\pm\frac{(\epsilon_\bk-\lambda)^2}{\sqrt{\alpha_\bk^2-\gamma_\bk^2}}\right)=\frac{4}{J_2}+4P\delta\Omega_{\textrm{2ch}}.
\end{split}
\label{MFEQNS}
\end{equation}
In the normal phase either $v_1$ or $v_2$ is nonzero, corresponding to
the development of the Kondo effect in the strongest channel. Here we will consider two cases: the first one corresponds
to the choice of $J_1/J_2$ such that resonances in both channels develop simultaneously, while in the second case the
condensation occurs in the first channel. From the mean-field equations (\ref{MFEQNS}) we can already see that the pressure has an opposite effect on the condensation temperatures in two channels: since $\delta\Omega_{\textrm{1ch}}<0$ it means that the effective exchange coupling $\tilde{J}_{1}(P)>J_1(P=0)$, signaling an increase of the Kondo lattice coherence temperature. In contrast, $\tilde{J}_2(P)<J_2(P=0)$, implying a decrease in $T_c$ with pressure. We compute the corresponding 
dependences of $T_c$ and $T_{\textrm{coh}}$ on pressure by solving Eqs. (\ref{MFEQNS}) numerically. The results are shown on Fig.~\ref{TcTcohP}. We note also that while opposite tendencies in $T_{\textrm{coh}}$ and $T_c$ in response to pressure appears to be a universal feature for the composite pairing mechanism, the rates with which $T_{\textrm{coh}}$ and $T_c$ change with pressure are not universal and depend on the microscopic features of the model.

\section{Summary}
Our findings on the Ce$_{1-x}$Yb$_x$CoIn$_5$ system are in line with the emerging scenario of two coexisting electronic networks (one of Ce $f$-moments and another of Yb $f$-electrons in an intermediate valence state) coupled to the conduction electrons. This is consistent with our observation that the quantum fluctuations are suppressed for $x = 0.20$ Yb doping (see $H_{QCP}$ in Fig. 1) but $T_c$ stays robust suggesting that spin fluctuations could not be the glue for SC pairing. Furthermore, the robust nature of $T_{\textrm{coh}}$ and $T_c$ and their scaling suggest that the emergence of SC and the onset of many-body coherence in the Kondo lattice have the same physical origin: hybridization between conduction and localized Ce $f$-electron states. 

We have established that within the mean-field theory approach for the composite pairing scenario, $T_c$ decreases while $T_{\textrm{coh}}$ increases with increasing $P$. These results are at odds with the experimental data in Ce$_{1-x}$Yb$_x$CoIn$_5$, which show that both $T_{\textrm{coh}}$ and $T_c$ initially increase with pressure (Fig.~5). By examining $T_c\sim T_{\textrm{coh}}\exp\left(-1/\nu_FJ_2\right)$ we see that one possible way to reconcile the composite pairing theory \cite{Flint2010,Erten2014} with the experimental data would be to assume that quantum valence fluctuations may effectively renormalize the spin exchange coupling $J_2$ to compensate for the effect caused by the change in the ionic volume. In this case, only $T_{\textrm{coh}}(P)$ will have a pronounced pressure dependence and $T_c(P)\sim T_{\textrm{coh}}(P)$, in agreement with experimental observations. However, a detailed study of this problem clearly goes beyond the scope of this paper and will be addressed elsewhere.

\ack 
This work was supported by the National Science Foundation (grant NSF DMR-1006606) and Ohio Board of Regents (grant OBR-RIP-220573) at KSU, and by the U.S. Department of Energy (grant DE-FG02-04ER46105) at UCSD. 

\section*{References}

\bibliography{Myref}

\begin{thebibliography}{10}

\bibitem{Norman2011}
M.~R. Norman {\em Science}, vol.~332, no.~6026, pp.~196--200, 2011.

\bibitem{Pfleiderer2009}
C.~Pfleiderer {\em Rev. Mod. Phys.}, vol.~81, pp.~1551--1624, Nov 2009.

\bibitem{Scalapino2012}
D.~J. Scalapino {\em Rev. Mod. Phys.}, vol.~84, pp.~1383--1417, Oct 2012.

\bibitem{Paglione2007}
J.~Paglione, T.~A. Sayles, P.~C. Ho, J.~R. Jeffries, and M.~B. Maple {\em Nat.
  Phys.}, vol.~3, pp.~703--706, September 2007.

\bibitem{Curro2005}
N.~Curro, T.~Caldwell, E.~Bauer, L.~Morales, M.~Graf, Y.~Bang, A.~Balatsky,
  J.~Thompson, and J.~Sarrao {\em Nature}, vol.~434, no.~7033, pp.~622--625,
  2005.

\bibitem{Sakai2005}
H.~Sakai, Y.~Tokunaga, T.~Fujimoto, S.~Kambe, R.~E.~Walstedt, H.~Yasuoka,
  D.~Aoki, Y.~Homma, E.~Yamamoto, A.~Nakamura, Y.~Shiokawa, K.~Nakajima,
  Y.~Arai, T.~D.~Matsuda, Y.~Haga, and Y.~Ōnuki {\em Journal of the Physical
  Society of Japan}, vol.~74, no.~6, pp.~1710--1713, 2005.

\bibitem{Bauer2012}
E.~D. Bauer, M.~M. Altarawneh, P.~H. Tobash, K.~Gofryk, O.~E. Ayala-Valenzuela,
  J.~N. Mitchell, R.~D. McDonald, C.~H. Mielke, F.~Ronning, J.-C. Griveau,
  E.~Colineau, R.~Eloirdi, R.~Caciuffo, B.~L. Scott, O.~Janka, S.~M.
  Kauzlarich, and J.~D. Thompson {\em Journal of Physics: Condensed Matter},
  vol.~24, no.~5, p.~052206, 2012.

\bibitem{Aoki2007}
D.~Aoki, Y.~Haga, T.~D.~Matsuda, N.~Tateiwa, S.~Ikeda, Y.~Homma, H.~Sakai,
  Y.~Shiokawa, E.~Yamamoto, A.~Nakamura, R.~Settai, and Y.~Ōnuki {\em Journal
  of the Physical Society of Japan}, vol.~76, no.~6, p.~063701, 2007.

\bibitem{Borisenko2010}
S.~V. Borisenko, V.~B. Zabolotnyy, D.~V. Evtushinsky, T.~K. Kim, I.~V. Morozov,
  A.~N. Yaresko, A.~A. Kordyuk, G.~Behr, A.~Vasiliev, R.~Follath, and
  B.~B\"uchner {\em Phys. Rev. Lett.}, vol.~105, p.~067002, Aug 2010.

\bibitem{Hu2012}
T.~Hu, H.~Xiao, T.~A. Sayles, M.~Dzero, M.~B. Maple, and C.~C. Almasan {\em
  Phys. Rev. Lett.}, vol.~108, p.~056401, Jan 2012.

\bibitem{Shu2011}
L.~Shu, R.~E. Baumbach, M.~Janoschek, E.~Gonzales, K.~Huang, T.~A. Sayles,
  J.~Paglione, J.~O'Brien, J.~J. Hamlin, D.~A. Zocco, P.-C. Ho, C.~A. McElroy,
  and M.~B. Maple {\em Phys. Rev. Lett.}, vol.~106, p.~156403, Apr 2011.

\bibitem{Hu2013}
T.~Hu, Y.~P. Singh, L.~Shu, M.~Janoschek, M.~Dzero, M.~B. Maple, and C.~C.
  Almasan {\em Proceedings of the National Academy of Sciences}, vol.~110,
  no.~18, pp.~7160--7164, 2013.

\bibitem{Singh2014}
Y.~P. Singh, D.~J. Haney, X.~Y. Huang, I.~K. Lum, B.~D. White, M.~Dzero, M.~B.
  Maple, and C.~C. Almasan {\em Phys. Rev. B}, vol.~89, p.~115106, Mar 2014.

\bibitem{Paglione2003}
J.~Paglione, M.~A. Tanatar, D.~G. Hawthorn, E.~Boaknin, R.~W. Hill, F.~Ronning,
  M.~Sutherland, L.~Taillefer, C.~Petrovic, and P.~C. Canfield {\em Phys. Rev.
  Lett.}, vol.~91, p.~246405, Dec 2003.

\bibitem{Sidorov2002}
V.~A. Sidorov, M.~Nicklas, P.~G. Pagliuso, J.~L. Sarrao, Y.~Bang, A.~V.
  Balatsky, and J.~D. Thompson {\em Phys. Rev. Lett.}, vol.~89, p.~157004, Sep
  2002.

\bibitem{Nair2010}
S.~Nair, O.~Stockert, U.~Witte, M.~Nicklas, R.~Schedler, K.~Kiefer, J.~D.
  Thompson, A.~D. Bianchi, Z.~Fisk, S.~Wirth, and F.~Steglich {\em Proceedings
  of the National Academy of Sciences}, 2010.

\bibitem{Zaum2011}
S.~Zaum, K.~Grube, R.~Sch\"afer, E.~D. Bauer, J.~D. Thompson, and
  H.~v.~L\"ohneysen {\em Phys. Rev. Lett.}, vol.~106, p.~087003, Feb 2011.

\bibitem{Jang2014}
S.~Jang, B.~White, I.~Lum, H.~Kim, M.~Tanatar, W.~Straszheim, R.~Prozorov,
  T.~Keiber, F.~Bridges, L.~Shu, {\em et~al.} {\em arXiv preprint
  arXiv:1407.6725}, 2014.

\bibitem{Petrovic2002}
C.~Petrovic, S.~Bud’ko, V.~Kogan, and P.~Canfield {\em Physical Review B},
  vol.~66, no.~5, p.~054534, 2002.

\bibitem{Dudy2013}
L.~Dudy, J.~D. Denlinger, L.~Shu, M.~Janoschek, J.~W. Allen, and M.~B. Maple
  {\em Phys. Rev. B}, vol.~88, p.~165118, Oct 2013.

\bibitem{Kim2014}
H.~Kim, M.~A. Tanatar, R.~Flint, C.~Petrovic, R.~Hu, B.~D. White, I.~K. Lum,
  M.~B. Maple, and R.~Prozorov {\em arXiv:1404.3700v1}, April 2014.

\bibitem{Dzero2012}
M.~Dzero and X.~Huang {\em Journal of Physics: Condensed Matter}, vol.~24,
  no.~7, p.~075603, 2012.

\bibitem{Polyakov2012}
A.~Polyakov, O.~Ignatchik, B.~Bergk, K.~G\"otze, A.~D. Bianchi, S.~Blackburn,
  B.~Pr\'evost, G.~Seyfarth, M.~C\^ot\'e, D.~Hurt, C.~Capan, Z.~Fisk, R.~G.
  Goodrich, I.~Sheikin, M.~Richter, and J.~Wosnitza {\em Phys. Rev. B},
  vol.~85, p.~245119, Jun 2012.

\bibitem{Jaccard1999}
D.~Jaccard, K.~Behnia, and J.~Sierro {\em Physics Letters A}, vol.~163,
  no.~5–6, pp.~475 -- 480, 1992.

\bibitem{Grosche1996}
F.~Grosche, S.~Julian, N.~Mathur, and G.~Lonzarich {\em Physica B: Condensed
  Matter}, vol.~223–224, no.~0, pp.~50 -- 52, 1996.
\newblock Proceedings of the International Conference on Strongly Correlated
  Electron Systems.

\bibitem{Almasan2014}
Y.~P. Singh, D.~J. Haney, X.~Huang, M.~B. Maple, M.~Dzero, and C.~C. Almasan
  {\em pre-print}, 2014.

\bibitem{Coleman1999}
P.~Coleman, A.~M. Tsvelik, N.~Andrei, and H.~Y. Kee {\em Phys. Rev. B},
  vol.~60, pp.~3608--3628, Aug 1999.

\bibitem{Flint2008}
R.~Flint, M.~Dzero, and P.~Coleman {\em Nat Phys}, vol.~4, pp.~643--648, 08
  2008.

\bibitem{Flint2010}
R.~Flint and P.~Coleman {\em Phys. Rev. Lett.}, vol.~105, p.~246404, Dec 2010.

\bibitem{Flint2011}
R.~Flint, A.~H. Nevidomskyy, and P.~Coleman {\em Phys. Rev. B}, vol.~84,
  p.~064514, Aug 2011.

\bibitem{Cox1997}
T.-S. Kim and D.~L. Cox {\em Phys. Rev. B}, vol.~55, pp.~12594--12619, May
  1997.

\bibitem{Coleman1993}
P.~Coleman, E.~Miranda, and A.~Tsvelik {\em Phys. Rev. Lett.}, vol.~70,
  pp.~2960--2963, May 1993.

\bibitem{Coleman1994}
P.~Coleman, E.~Miranda, and A.~Tsvelik {\em Phys. Rev. B}, vol.~49,
  pp.~8955--8982, Apr 1994.

\bibitem{Hoshino2014}
S.~Hoshino and Y.~Kuramoto {\em Phys. Rev. Lett.}, vol.~112, p.~167204, Apr
  2014.

\bibitem{Anders2002PRB}
F.~B. Anders {\em Phys. Rev. B}, vol.~66, p.~020504, Jul 2002.

\bibitem{Anders2002EPJ}
F.~Anders {\em The European Physical Journal B - Condensed Matter and Complex
  Systems}, vol.~28, no.~1, pp.~9--28, 2002.

\bibitem{Zhang2002}
S.~Zhang {\em Phys. Rev. B}, vol.~65, p.~064407, Jan 2002.

\bibitem{Erten2014}
O.~Erten, R.~Flint, and P.~Coleman {\em arXiv:1402.7361v2}, Feb 2014.

\end{thebibliography}

\end{document}